\documentclass{webofc}

\usepackage[varg]{txfonts}   
\usepackage{hyperref}

\usepackage{url}

\hypersetup{colorlinks=true,citecolor=blue,urlcolor=blue,linkcolor=blue}
\begin{document}
\begin{flushright}
OU-HET-1248
\end{flushright}

\title{New renormalization scheme in extended Higgs sectors for Higgs precision measurements}

%
%

\author{\firstname{Shinya} \lastname{Kanemura}\inst{1}\fnsep\thanks{\email{kanemu@het.phys.sci.osaka-u.ac.jp}} \and
        \firstname{Mariko} \lastname{Kikuchi}\inst{2}\fnsep\thanks{\email{kikuchi.mariko13@nihon-u.ac.jp}} \and
        \firstname{Kei} \lastname{Yagyu}\inst{1}\fnsep\thanks{\email{yagyu@het.phys.sci.osaka-u.ac.jp}}\footnote[5]{Speaker}
}

\institute{Department of Physics, Osaka University, Toyonaka, Osaka 560-0043, Japan 
\and
           College of Engineering, Nihon University, Koriyama, Fukushima 963-8642, Japan
          }

\abstract{We discuss a new renormalization scheme for mixing angles in extended Higgs sectors for the coming era of the Higgs precise measurements at future lepton colliders. 
We focus on the two Higgs doublet models (2HDMs) with a softly-broken $Z_2$ symmetry as a simple and important example, in which two mixing angles $\alpha$ and $\beta$ appear in the Higgs sector.   
In this new scheme, the counterterms for two mixing angles $\delta\alpha$ and $\delta\beta$ are determined by requiring that deviations in the decay rates of $h\to ZZ^* \to Z\ell^+\ell^-$ and $h \to \tau\tau$ 
from the corresponding predictions in the standard model at NLO are given by the square of the scaling factor at tree level. 
We show how this scheme works in the 2HDMs, and demonstrate how the other decay rates (e.g., $h \to WW^*$, $h \to b\bar{b}$, etc.) are predicted at NLO.
}
\maketitle
\section{Introduction}
\label{intro}

Precise measurements of the properties of the discovered Higgs boson provide a quite important and robust way 
to indirectly search for new physics beyond the Standard Model (SM). 
Namely, effects of new physics can appear via deviations in Higgs observables from the SM predictions, e.g., production cross sections, decay branching ratios, 
and their patterns and magnitudes strongly depend on new physics scenarios. 
Such deviations can be observed at future Higgs factories, e.g., 
the International Linear Collider (ILC)~\cite{Baer:2013cma,Asai:2017pwp,Fujii:2017vwa,ILC:2019gyn}, the Circular Electron-Positron Collider (CEPC)~\cite{CEPC-SPPCStudyGroup:2015csa}, the Future Circular Collider (FCC-ee)~\cite{FCC:2018byv} and the Compact LInear Collider (CLIC)~\cite{Klamka:2021cjt}.
Therefore, it is necessary to calculate the Higgs properties with an accuracy equal to or better than that expected in these experiments, typically better than 1\% level, 
so that radiative corrections have to be included in the calculation. 

There are basically two sources which induce deviations in the Higgs properties, i.e., scalar mixings at tree level and radiative corrections. 
The former can simply be described in terms of the so-called scaling factors $\kappa_X^{}$ which represent the value of the Higgs boson coupling normalized by the corresponding SM prediction, and 
are expressed by the mixing angles. 
Thus, the mixing angles play a role to describe the ``alignmentness'' at tree level, i.e., how the Higgs properties are close to those of the SM Higgs boson. 
The situation, however, can drastically be changed when we consider radiative corrections to the Higgs properties, because other model parameters can also enter into the predictions.  
Therefore, the mixing parameters no longer describe the alignmentness at loop levels. 

In this presentation, we propose a new renormalization scheme for the mixing parameters such that they still describe the alignmentness at loop levels, based on the recent paper~\cite{Kanemura:2024ium}. 
As a simple and important example, we discuss the two Higgs doublet models (2HDMs) with a softly-broken $Z_2$ symmetry in order to show how our scheme works in a concrete way.

\section{2HDMs}

We briefly review the 2HDMs with a softly-broken $Z_2$ symmetry to avoid flavor changing neutral currents at tree level, 
under which two doublets $\Phi_1$ and $\Phi_2$ are transformed as $(\Phi_1,\Phi_2) \to (\Phi_1,-\Phi_2)$. 

The Higgs potential is generally given by 
\begin{align}
V&= m_1^2|\Phi_1|^2 + m_2^2|\Phi_2|^2 - m_3^2(\Phi_1^\dagger \Phi_2 +\text{h.c.}) \notag\\
&+\frac{\lambda_1}{2}|\Phi_1|^4+\frac{\lambda_2}{2}|\Phi_2|^4 + \lambda_3|\Phi_1|^2|\Phi_2|^2 + \lambda_4|\Phi_1^\dagger\Phi_2|^2
+ \frac{\lambda_5}{2}\left[(\Phi_1^\dagger\Phi_2)^2+\text{h.c.}\right], \label{pot_thdm1}
\end{align}
where $m_3^2$ and $\lambda_5$ are assumed to be real for simplicity. 
We introduce the Higgs basis as 
\begin{align}
&\left(\begin{array}{c}
\Phi_1\\
\Phi_2
\end{array}\right)=
\left(\begin{array}{cc}
\cos\beta & -\sin\beta\\
\sin\beta & \cos\beta
\end{array}\right)
\left(\begin{array}{c}
\Phi\\
\Phi'
\end{array}\right), 
\end{align}
where $\tan\beta = v_2/v_1$ with $v_a = \sqrt{2}\langle \Phi_a^0\rangle$, and 
\begin{align}
\Phi=\left[
\begin{array}{c}
G^+\\
\frac{1}{\sqrt{2}}(h_1'+v+ iG^0)
\end{array}\right],\quad
\Phi'=\left[
\begin{array}{c}
H^+\\
\frac{1}{\sqrt{2}}(h_2'+iA)
\end{array}\right]. \label{Higgs-basis}
\end{align}
In the above expression, $H^\pm$, $A$ and $h_a^\prime$ ($a = 1,2$) respectively represent
the physical singly-charged, CP-odd and CP-even Higgs bosons, while 
$G^\pm$ ($G^0$) are the Nambu-Goldstone (NG) bosons absorbed into the longitudinal components of the $W^\pm$ ($Z$) bosons.  
The Vacuum Expectation Value (VEV) $v$ is related to the Fermi constant $G_F$ via $v\equiv \sqrt{v_1^2+v_2^2}=(\sqrt{2}G_F)^{-1/2}\simeq 246$ GeV.
The two CP-even Higgs bosons are mixed with each other, and their mass eigenstates can be defined as 
\begin{align}
\begin{pmatrix}
h_1^\prime \\
h_2^\prime
\end{pmatrix} =
\begin{pmatrix}
\cos(\beta-\alpha) & \sin(\beta-\alpha) \\
-\sin(\beta-\alpha) & \cos(\beta-\alpha)
\end{pmatrix}
\begin{pmatrix}
H\\
h
\end{pmatrix}, 
\end{align}
with $h$ being identified with the discovered Higgs boson. 

After solving the stationary conditions, we obtain the 
masses of physical Higgs bosons as 
\begin{align}
m_{H^\pm}^2=M^2-\frac{v^2}{2}(\lambda_4+\lambda_5),\quad m_A^2&=M^2-v^2\lambda_5, \label{mass1}
\end{align}
where $M^2\equiv m_3^2/(\sin\beta\cos\beta)$, and 
the squared mass matrix for the CP-even Higgs bosons is given in the Higgs basis ($h_1^\prime,h_2^\prime$) as 
\begin{align}
{\cal M} =v^2 
\begin{pmatrix}
\lambda_1c^4_\beta + \lambda_2 s^4_\beta +\frac{\lambda_{345}}{2}s^2_{2\beta}
& \frac{1}{2}(\lambda_2 s^2_\beta -\lambda_1c^2_\beta + \lambda_{345}c_{2\beta}) s_{2\beta}\\
\frac{1}{2}(\lambda_2 s^2_\beta -\lambda_1c^2_\beta + \lambda_{345}c_{2\beta}) s_{2\beta}& \frac{M^2}{v^2} + \frac{1}{4}(\lambda_1+\lambda_2-2\lambda_{345}) s^2_{2\beta}
\end{pmatrix},  \label{eq:lams}
\end{align}
where $\lambda_{345}\equiv \lambda_3+\lambda_4+\lambda_5$ with $c_\theta \equiv \cos\theta$, $s_\theta \equiv \sin\theta$ and $t_\theta \equiv \tan\theta$. 
The mass eigenvalues and the mixing angle are then expressed by 
\begin{align}
m_H^2&= {\cal M}_{11}\, c^2_{\beta-\alpha}+{\cal M}_{22}\, s^2_{\beta-\alpha} -{\cal M}_{12}\, s_{2(\beta-\alpha)},\\
m_h^2&= {\cal M}_{11}\,  s^2_{\beta-\alpha}+{\cal M}_{22}\,  c^2_{\beta-\alpha}+{\cal M}_{12}\, s_{2(\beta-\alpha)},\\
\tan 2(\beta-\alpha)&=\frac{2{\cal M}_{12}}{{\cal M}_{22}-{\cal M}_{11}}. \label{Eq:tan2}
\end{align} 
It is clear that the limit $M^2 \to \infty$ corresponds to the decoupling limit, where all the masses of the additional Higgs bosons become infinity, and only the $h$ state stays at the electroweak scale. 
In this limit, the alignment limit $s_{\beta-\alpha} \to 1$ is also realized, because the off-diagonal element of the mass matrix (\ref{eq:lams}), determined by $v^2$, is negligibly small as compared with the (2,2) element. 
We note that inverse of this statement is not true in general, because the alignment limit $s_{\beta-\alpha} \to 1$ can be taken by choosing the off-diagonal element of (\ref{eq:lams}) to be very small regardless the value of $M^2$. 
In particular, the so-called alignment without decoupling, $M^2 \simeq {\cal O}(v^2)$ with $s_{\beta-\alpha} \simeq 1$ are well motivated for various new physics scenarios such as the electroweak baryogenesis, see e.g.,~\cite{Enomoto:2021dkl,Enomoto:2022rrl,Kanemura:2023juv}. 

The Yukawa interactions are expressed in the Higgs basis as 
\begin{align}
{\cal L}_{\rm Y} &= -
\frac{\sqrt{2}}{v}\left[
\bar{Q}_L M_u (\tilde{\Phi} + \zeta_u \tilde{\Phi}')u_R 
+\bar{Q}_L M_d(\Phi + \zeta_d \Phi')d_R 
+\bar{L}_L M_e(\Phi + \zeta_e \Phi')e_R\right] + \text{h.c.},  
\end{align}
where $M_f$ ($f=u,d,e$) are the diagonalized mass matrices for fermions, 
$\tilde{\Phi}^{(\prime)} = i\tau_2\Phi^{(\prime)*}$ and $\zeta_f$ are the flavor universal parameters depending on the four types of Yukawa interactions~\cite{Barger:1989fj,Grossman:1994jb,Aoki:2009ha} as shown in Tab.~\ref{tab-1}. 
%

\begin{table}
\centering
\caption{$\zeta_f$ factors in different types of the Yukawa interaction. }
\label{tab-1}       
\begin{tabular}{llll}
\hline
&$\zeta_u$ & $\zeta_d$ & $\zeta_e$  \\\hline
Type-I& $\cot\beta$ & $\cot\beta$ & $\cot\beta$ \\
Type-II&$\cot\beta$ & $-\tan\beta$ & $-\tan\beta$ \\
Type-X (Lepton specific)&$\cot\beta$ & $\cot\beta$ & $-\tan\beta$ \\
Type-Y (Flipped)&$\cot\beta$ & $-\tan\beta$ & $\cot\beta$ \\\hline
\end{tabular}
\end{table}

\section{New renormalization scheme}

We here discuss the essence of our new scheme. See Ref.~\cite{Kanemura:2024ium} for more detailed discussions. 

We first shift all the parameters in the Higgs potential as 
\begin{align}
\begin{split}
& v \to v + \delta v, \\
&m_\varphi^2 \to m_\varphi^2 + \delta m_\varphi^2 \quad (\varphi = h,~H,~A,~H^\pm), \\
& M^2 \to M^2 + \delta M^2, \\
& \alpha \to \alpha + \delta \alpha,\quad 
\beta \to \beta + \delta \beta. 
\end{split} \label{eq:1}
\end{align}
The tadpole counterterm can also be introduced, and we here apply the alternative tadpole scheme~\cite{Fleischer:1980ub,Krause:2016oke}, i.e., instead of introducing the tadpole counterterms, 
we add the tadpole inserted diagrams into the one-particle irreducible diagrams. 
Next, we shift the wavefunctions of the scalar states as 
\begin{align}
\begin{pmatrix}
H \\
h 
\end{pmatrix}
&\to 
Z_{\rm even}
\begin{pmatrix}
H \\
h 
\end{pmatrix}, \quad
\begin{pmatrix}
G^0 \\
A 
\end{pmatrix}
\to 
Z_{\rm odd}
\begin{pmatrix}
G^0 \\
A 
\end{pmatrix}, \quad
\begin{pmatrix}
G^\pm \\
H^\pm 
\end{pmatrix}
\to 
Z_\pm
\begin{pmatrix}
G^\pm \\
H^\pm 
\end{pmatrix}, \label{eq:2}
\end{align}
where $Z_{\rm even}$, $Z_{\rm odd}$ and $Z_\pm$ are $2\times 2$ matrices for the wavefunction renormalization, and each of them is expressed as 
$Z_{\rm even} = I_{2\times 2} + \frac{1}{2}\delta Z_{\rm even}$ and similar for the CP-odd and charged states. 
From Eqs.~(\ref{eq:1}) and (\ref{eq:2}), we introduced 20 counterterms. 

The renormalized scalar two-point functions are then expressed as 
\begin{align}
\hat{\Pi}_{ij}(p^2) &= \Pi_{ij}(p^2) + (p^2-m_i^2)\frac{\delta Z_{ij}}{2} + (p^2-m_j^2)\frac{\delta Z_{ji}}{2} - \delta_{ij}\delta m_i^2, 
\end{align}
where the indices $i$ and $j$ represent all the possible scalar states including NG bosons, and  
$\Pi_{ij}$ are the unrenormalized two-point functions for the external $i$--$j$ state ($\Pi_{ij} = \Pi_{ji}$ are satisfied for $i\neq j$, and thus $\hat{\Pi}_{ij} = \hat{\Pi}_{ji}$). 
There are three different $\hat{\Pi}_{ij}$ functions for each CP-even, CP-odd and charged scalar sectors, e.g., 
for the CP-even sector, we have $(i,j) = (H,H)$, $(h,h)$ and $(H,h)$, and $\delta Z_{ij}$ should be understood as $(\delta Z_{\rm even})_{ij}$. 
We note that the mass counterterms $\delta m_i^2$ are zero for the NG bosons. 

In order to determine these counterterms, we impose the following on-shell renormalization conditions: 
\begin{align}
& \hat{\Pi}_{ii}(m_i^2) = 0, \quad \frac{d\hat{\Pi}_{ii}(p^2)}{dp^2}\Big|_{p^2 = m_i^2} = 0, \quad 
 \hat{\Pi}_{ij}(m_i^2) = 0~~(i\neq j), 
\end{align}
where $m_i^2 = 0$ for the NG bosons. 
We note that the left equation for the NG bosons, i.e., $i=G^\pm$ and $G^0$ at $p^2 = 0 $ is automatically satisfied, so that it does not determine any counterterms. 
Thus, 16 counterterms are determined by imposing the above three conditions as follows: 
\begin{align}
\delta m_i^2 &= \Pi_{ii}(m_i^2)~~(i\neq G^\pm,~G^0),\quad
\delta Z_{ii}   = -\frac{d\Pi_{ii}(p^2)}{dp^2}\Big|_{p^2 = m_i^2},\quad \delta Z_{ij}   &= \frac{2}{m_i^2 - m_j^2}\Pi_{ij}(m_j^2). 
\end{align}
As we have introduced 20 counterterms, 4 counterterms are not determined at this stage. 
We can determine $\delta v$ by using the renormalization in the electroweak sector as in the SM~\cite{Bohm:1986rj}. 
For $\delta M^2$, we can impose the $\overline{\text{MS}}$ scheme such that the ultra-violet divergent part of the $hhh$ vertex is cancelled at one-loop level~\cite{Kanemura:2004mg}. 

We still have two unrenormalized counterterms, i.e., $\delta\alpha$ and $\delta\beta$. 
In Ref.~\cite{Kanemura:2004mg} (we call it as the KOSY scheme), these mixing counterterms are determined by demanding 
\begin{align}
\delta \alpha = \frac{1}{4}(\delta Z_{hH} - \delta Z_{Hh}),\quad 
\delta \beta = \frac{1}{4}(\delta Z_{AG^0} - \delta Z_{G^0A}). 
\end{align}
These determinations are eventually the same as those proposed in Ref.~\cite{Pilaftsis:1997dr}
in models with scalar mixings, which have been obtained by the analogy to the method proposed by Denner and Sack~\cite{Denner:1990yz} for determining the quark-mixing matrix.\footnote{There have also been discussions of the 
$\overline{\rm MS}$ scheme for the determination of the mixing counterterms~\cite{Denner:2016etu,Denner:2017vms,Altenkamp:2017ldc,Denner:2018opp}.} 

We require the following renormalization conditions to determine $\delta\alpha$ and $\delta\beta$: 
\begin{align}
\Gamma(h \to Z\ell\bar{\ell})_{\rm NLO} &= (\kappa_V^{})^2\times\Gamma(h \to Z\ell\bar{\ell})_{\rm NLO}^{\rm SM}, \label{eq:4} \\
\Gamma(h\to \tau\bar{\tau})_{\rm NLO}&= (\kappa_\tau)^2\times \Gamma(h\to \tau\bar{\tau})_{\rm NLO}^{\rm SM}, \label{eq:3}
\end{align} 
where $\Gamma(h \to Z\ell\bar{\ell})_{\rm NLO}$ and $\Gamma(h\to \tau\bar{\tau})_{\rm NLO}$ are the decay rates of $h \to ZZ^*\to Z\ell^+\ell^-$ and $h \to \tau^+\tau^-$ at NLO in the 2HDMs, respectively. 
The $\kappa$ factors are given by 
\begin{align}
\kappa_V^{} = s_{\beta-\alpha},\quad 
\kappa_\tau = s_{\beta-\alpha} + \zeta_e c_{\beta-\alpha}, 
\end{align}
where $\zeta_e$ is given in Table~\ref{tab-1}. 
Writing these NLO decay rates as 
\begin{align}
\Gamma(h \to Z\ell\bar{\ell})_{\rm NLO} &= \Gamma(h \to Z\ell\bar{\ell})_{\rm LO}(1 + \Delta_{\rm EW}^{Z\ell\ell} ), \\
\Gamma(h\to \tau\bar{\tau})_{\rm NLO}&= \Gamma(h\to \tau\bar{\tau})_{\rm LO} (1 + \Delta_{\text{EW}}^\tau), 
\end{align} 
the conditions (\ref{eq:4}) and (\ref{eq:3}) are expressed as 
\begin{align}
\Delta_{\rm EW}^{Z\ell\ell} = \Delta_{\rm EW}^{Z\ell\ell}\Big|_{\rm SM},\quad 
\Delta_{\text{EW}}^\tau = \Delta_{\text{EW}}^\tau\Big|_{\rm SM},
\end{align} 
where $\Delta_{\rm EW}^{Z\ell\ell}$ and $\Delta_{\text{EW}}^\tau$ 
are the electroweak corrections to the decay rates of $h \to ZZ^* \to Z\ell^+\ell^-$ and $h \to \tau^+\tau^-$, respectively, 
and the former (latter) includes the counterterm $\delta (\beta-\alpha)$  ($\delta \beta$ and $\delta(\beta-\alpha)$) with $\delta (\beta-\alpha) \equiv \delta \beta-\delta\alpha$. 
See~\cite{Kanemura:2024ium} for the detailed expressions for $\delta(\beta-\alpha)$ and $\delta\beta$. 

\begin{figure}[!t]
\begin{center}
\includegraphics[width=60mm]{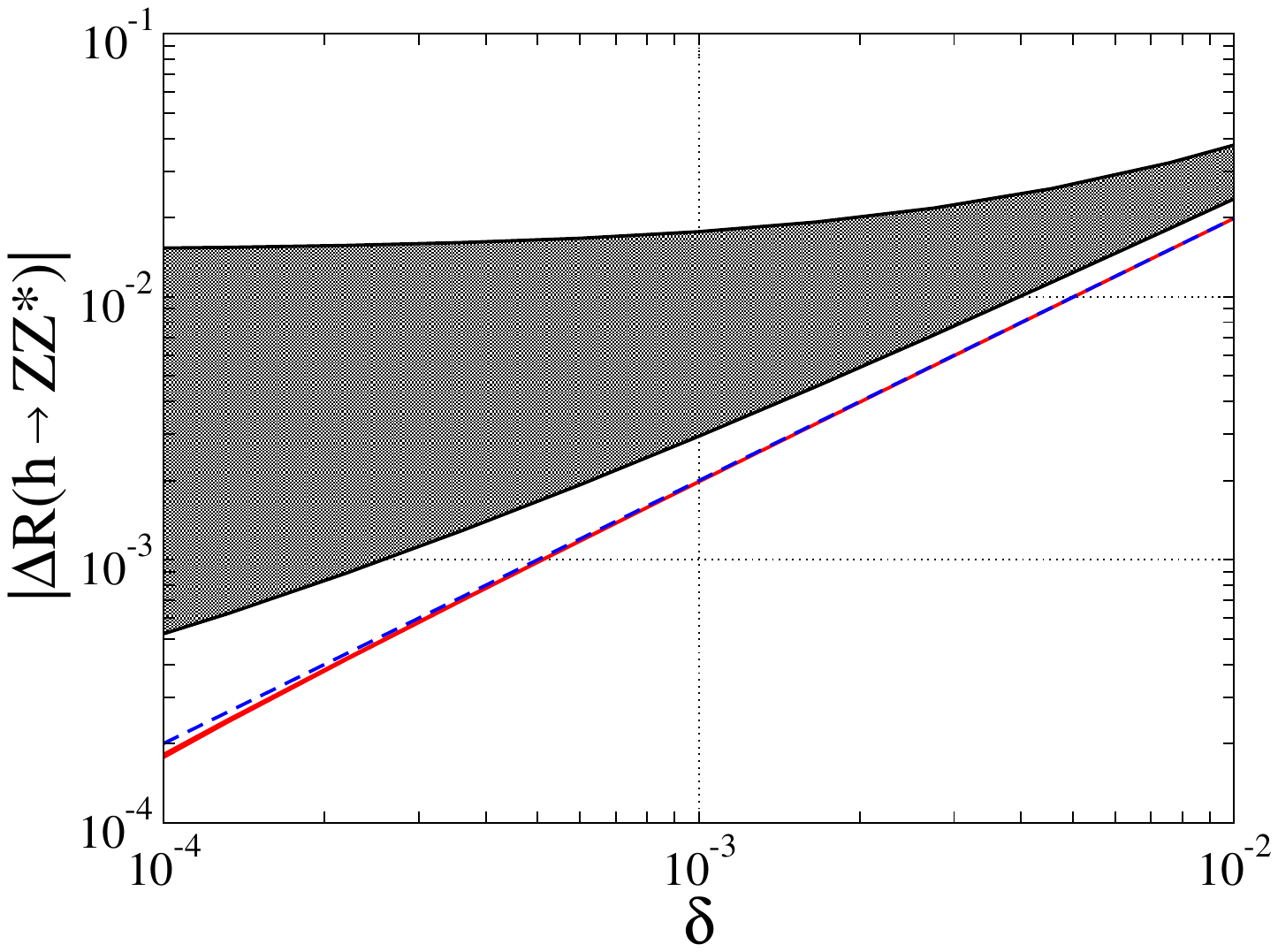}
\includegraphics[width=60mm]{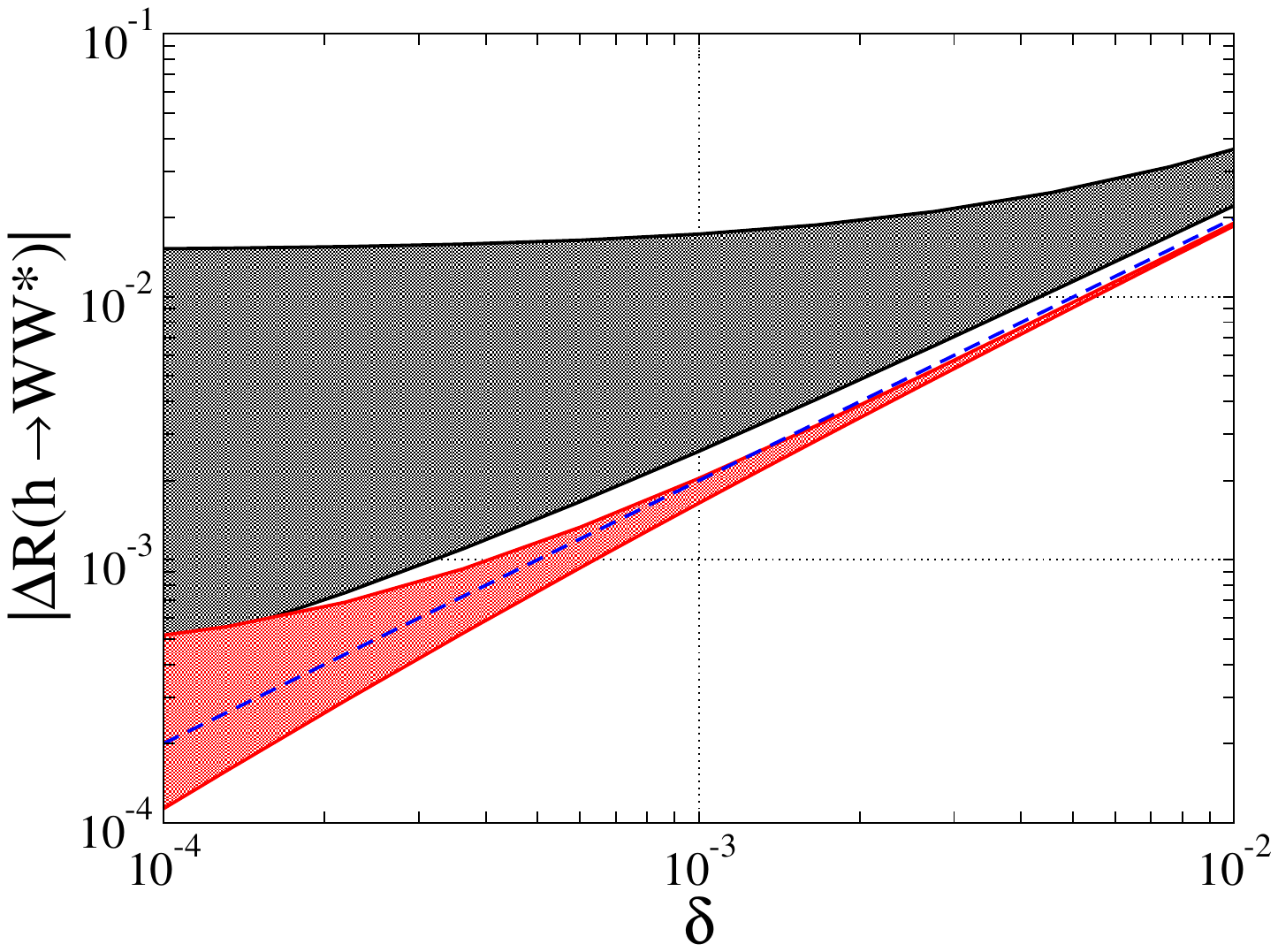}\\
\vspace{-0.1cm}
\includegraphics[width=60mm]{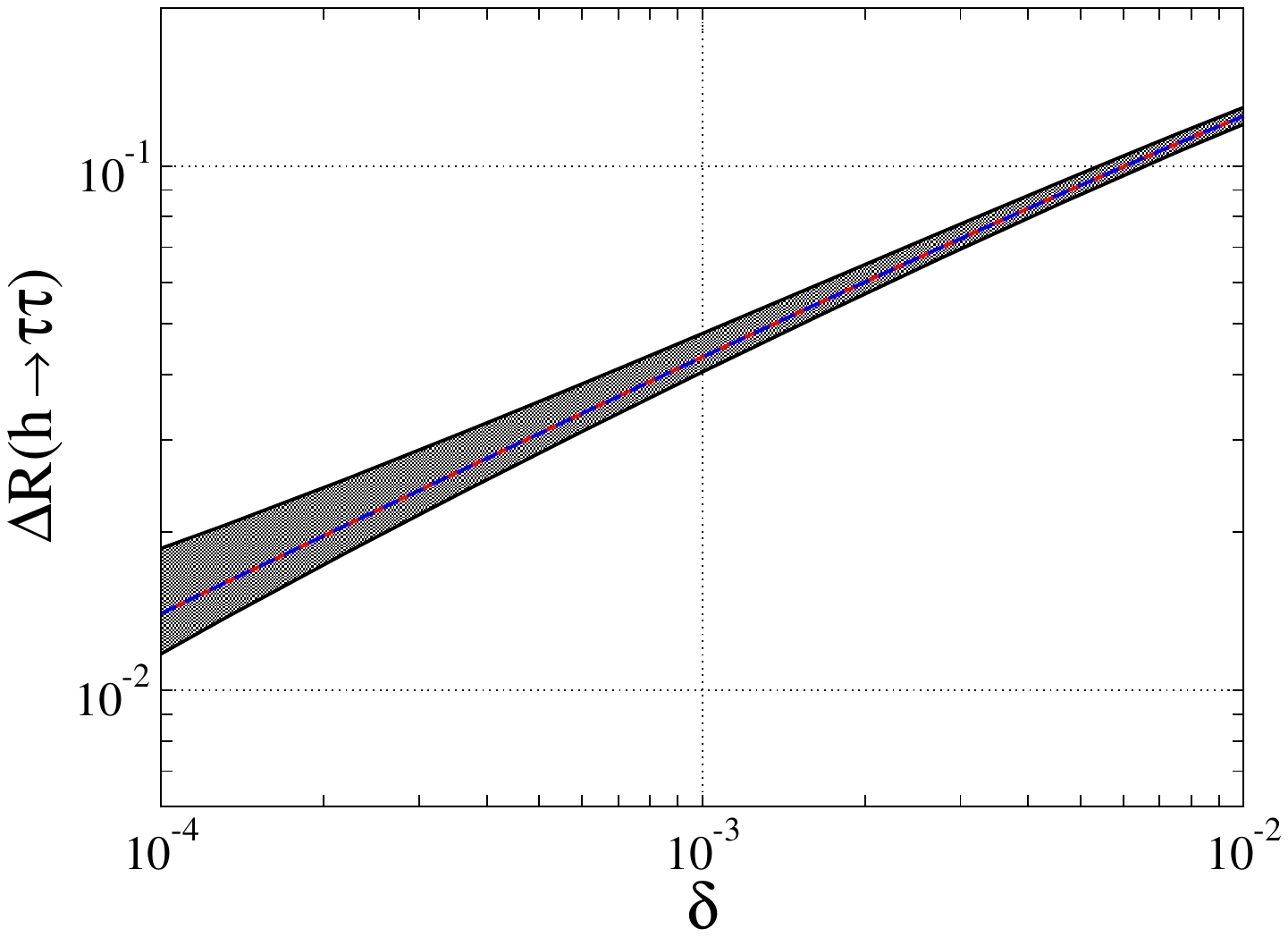}
\includegraphics[width=60mm]{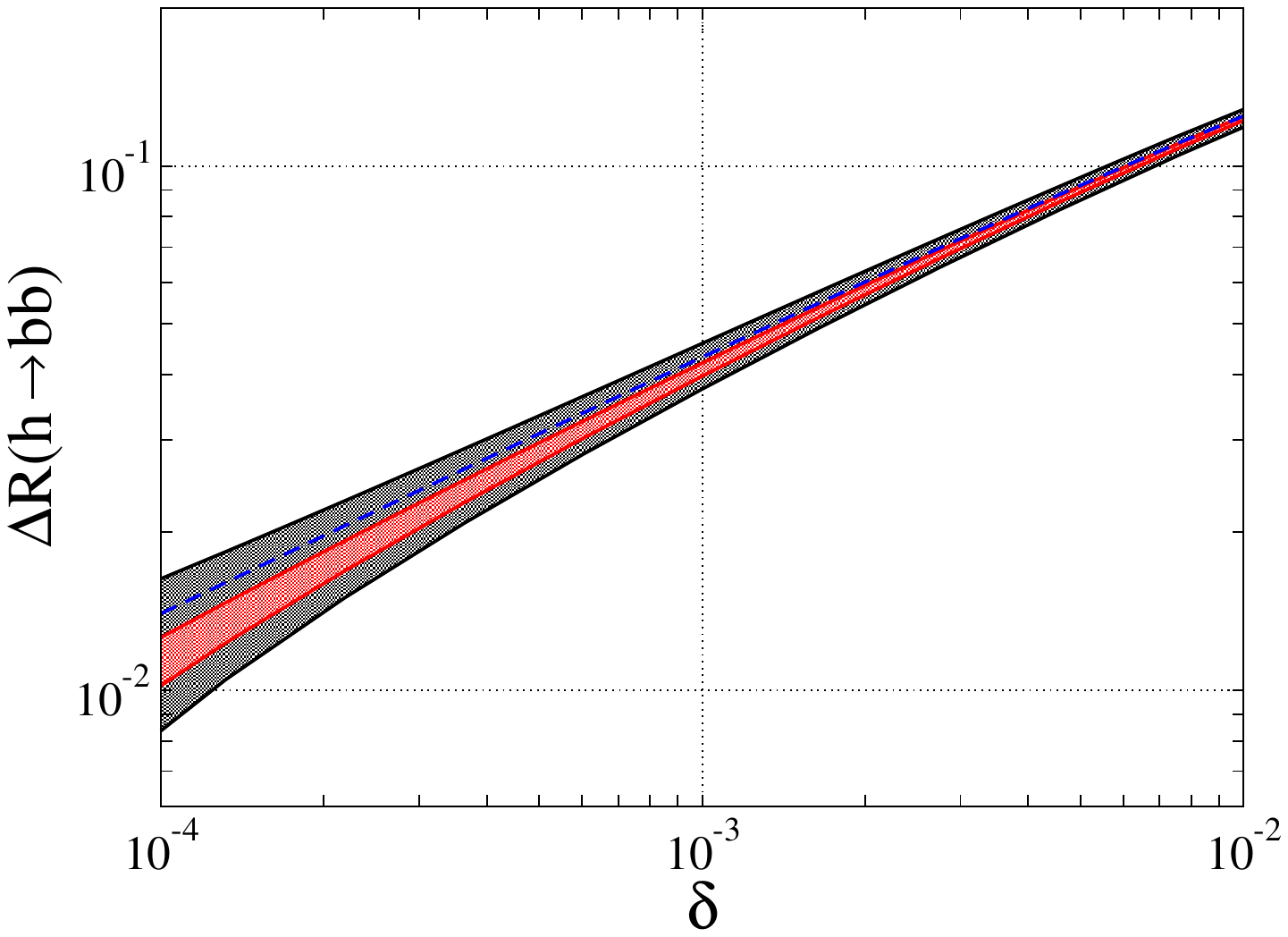}
\end{center}
\caption{Deviations in the decay rates $\Delta R$ as a function of $\delta (\equiv 1 - \sin(\beta-\alpha))$ in the Type-I 2HDM with $m_{H^\pm} = m_H = m_A = 300$ GeV, $\tan\beta = 2$, $\cos(\beta-\alpha) > 0$ and $M^2$ being scanned under the constraints from perturbative unitarity and vacuum stability.
The black and red shaded regions represent the predictions by using the KOSY scheme~\cite{Kanemura:2004mg} and the new scheme~\cite{Kanemura:2024ium}, respectively. 
The blue dashed curve show the LO result. }
\label{fig1}
\end{figure}

\begin{figure}[!t]
\begin{center}
\includegraphics[width=60mm]{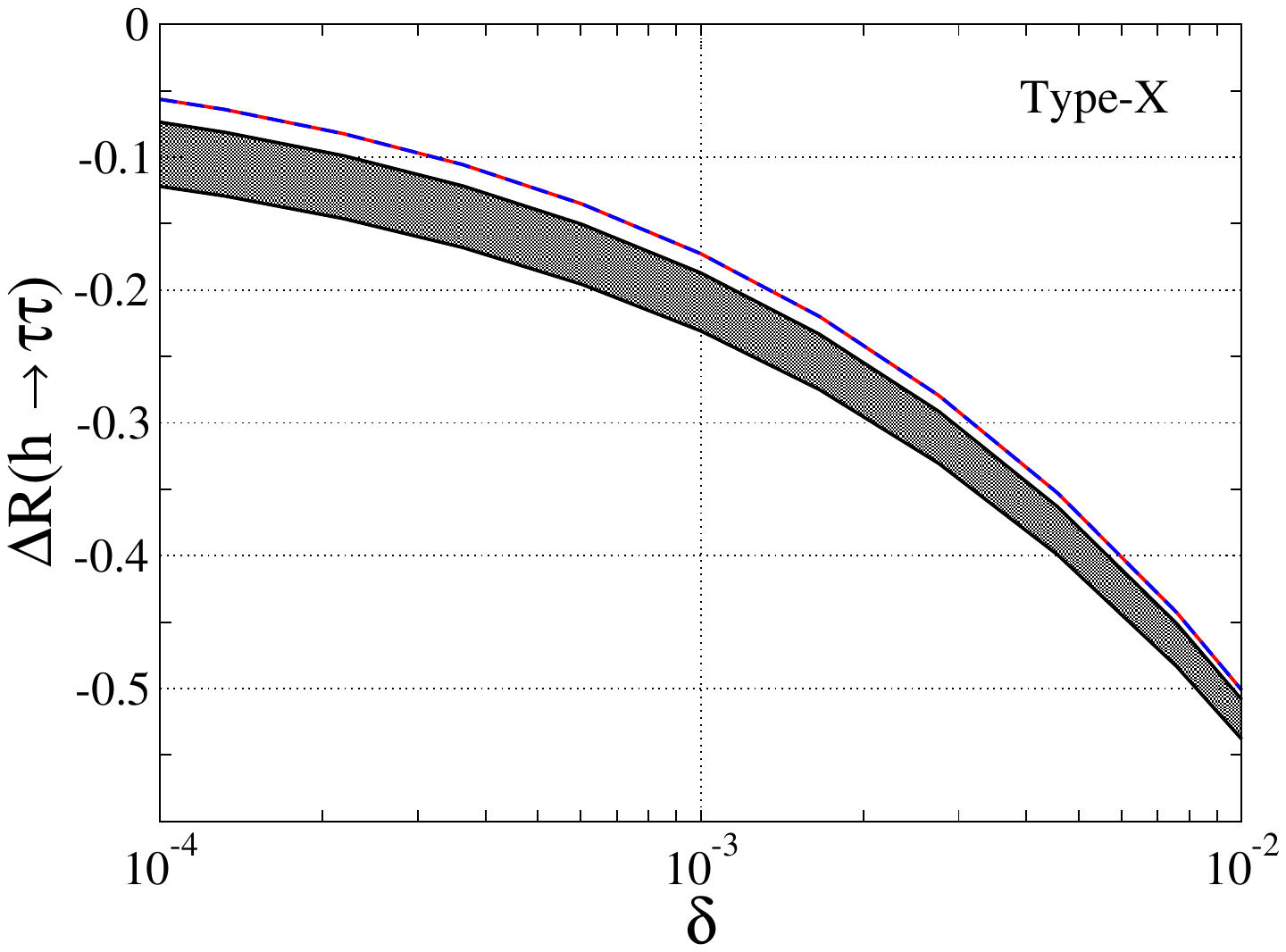}
\includegraphics[width=60mm]{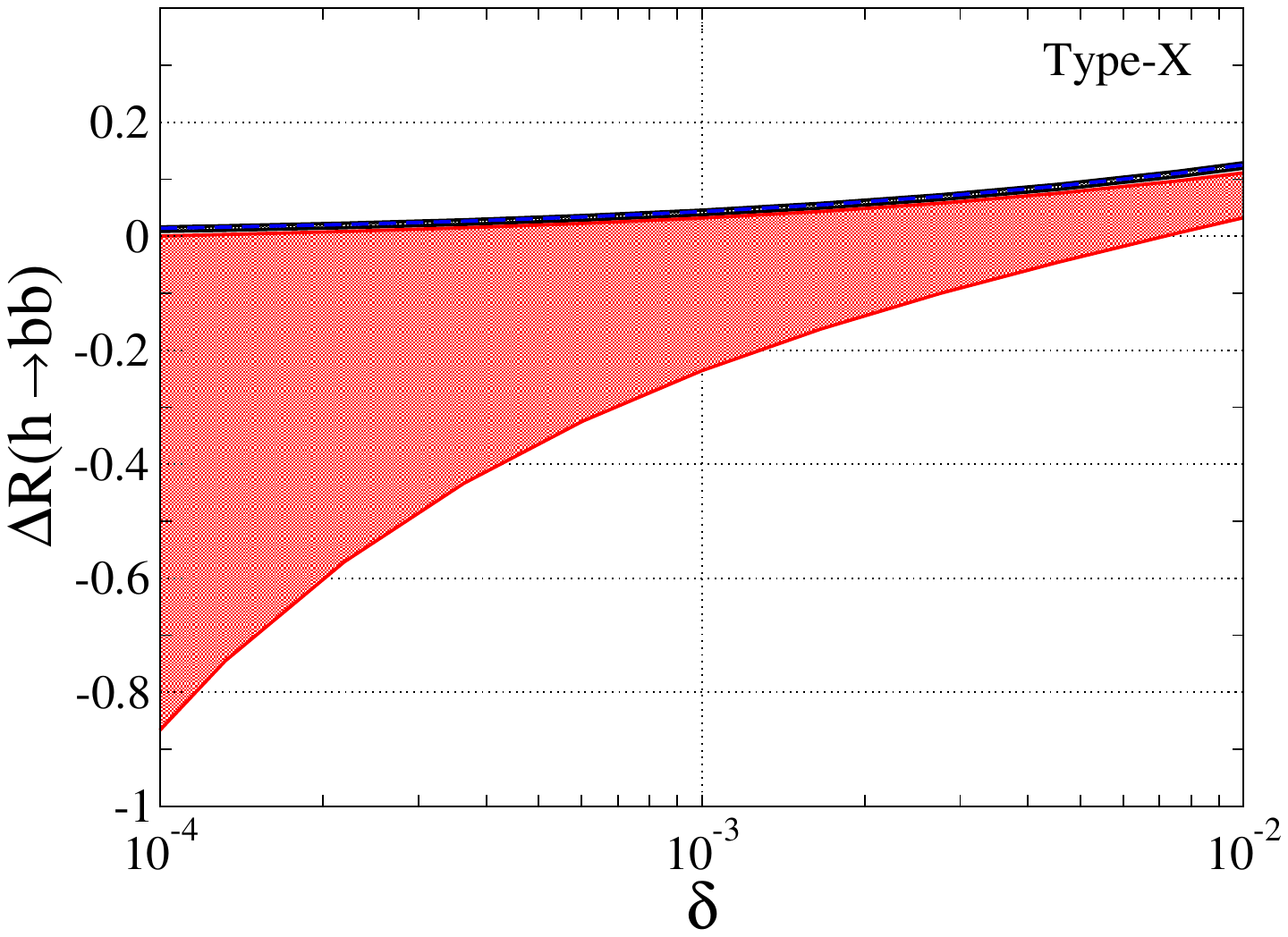}
\end{center}
\caption{Same as in Fig.~\ref{fig1}, but for $\Delta R(h \to \tau^+\tau^-)$ and $\Delta R(h \to b\bar{b})$ in the Type-X 2HDM. }
\label{fig2}
\end{figure}

Now, let us show numerical results for the decay rates at NLO by using the new scheme discussed above. 
We here employ the {\tt H-COUP} package~\cite{Kanemura:2017gbi,Kanemura:2019slf,Aiko:2023xui} for the numerical evaluation of the decay rates at NLO by implementing the new renormalization scheme into the latest version (Ver.~3).

In Fig.~\ref{fig1}, we show the deviation in the decay rates $\Delta R$ as a function of $\delta (\equiv 1 - s_{\beta-\alpha})$ in the Type-I 2HDM, where $\Delta R$ is defined as 
\begin{align}
\Delta R(h \to XY) \equiv  \frac{\Gamma(h \to XY)_{\rm NLO}}{\Gamma(h \to XY)_{\rm NLO}^{\rm SM}} - 1. \label{eq:delr}
\end{align}
We compare the results given in the KOSY scheme~\cite{Kanemura:2004mg}, the new scheme and the corresponding LO one. 
We here scan the value of $M^2$ under the constraints from the perturbative unitarity bound~\cite{Kanemura:1993hm,Akeroyd:2000wc,Ginzburg:2005dt,Kanemura:2015ska} and the vacuum stability bound~\cite{Deshpande:1977rw,Nie:1998yn,Kanemura:1999xf}.
It is seen that the results for $\Delta R(h \to ZZ^*)$ and $\Delta R(h \to \tau^+\tau^-)$
given in the new scheme and in the LO show quite good agreement with each other as these are taken as the inputs in the renormalization conditions. 
The results for  $\Delta R(h \to WW^*)$ and $\Delta R(h \to b\bar{b})$ can be regarded as the prediction in the new scheme, and they also show 
good agreement with the corresponding LO results. 

In Fig.~\ref{fig2}, we show $\Delta R(h \to \tau^+\tau^-)$ and $\Delta R(h \to b\bar{b})$ in the Type-X 2HDM. 
As in the Type-I 2HDM, $\Delta R(h \to \tau^+\tau^-)$ given in the new scheme shows quite good agreement with that given at LO. 
On the other hand, we see the large difference between $\Delta R(h \to b\bar{b})$ given in the new scheme and the LO result, especially in the case with smaller $\delta$. 
This can be explained by the contribution from the $\delta Z_h$ (wavefunction renormalization for $h$) term in $\Delta_{\rm EW}^\tau$ which is proportional to $\tan(\beta-\alpha)$ after substituting the expressions of 
$\delta\beta$ and $\delta(\beta-\alpha)$ determined by the renormalization conditions (\ref{eq:4}) and (\ref{eq:3}).
The counterterm $\delta Z_h$ typically gives ${\cal O}(1)\%$ corrections to the couplings of $h$ at one-loop level for $M^2/v^2 \ll 1$, and now its correction is enhanced by the factor of $\tan(\beta-\alpha)$ 
in the nearly alignment case, i.e., $c_{\beta-\alpha} \ll 1$. 
Such a large correction appears when $\zeta_\tau \neq \zeta_f$ ($f \neq \tau$), e.g., $\zeta_\tau = -\tan\beta$ and $\zeta_b = \cot\beta$ as in the Type-X 2HDM, 
while the $\delta Z_h$ dependence disappears in $\Delta_{\rm EW}^f$ for $\zeta_\tau= \zeta_f$ as in the Type-I case. 

\section{Conclusions}

We have discussed the new renormlization scheme for the mixing counterterms $\delta \alpha$ and $\delta \beta$ in the 2HDMs with a softly-broken $Z_2$ symmetry, in which 
these counterterms are determined by demanding that  the NLO decay rates of $h \to Z\ell^+\ell^-$ and $h \to \tau^+\tau^-$ are determined by the corresponding NLO prediction in the SM multiplied by 
the squared scaling factor at tree level. 
We have shown that the deviations in the decay rates of $h \to ZZ^*$ and $h \to WW^*$ from the SM predictions at NLO
are well described by $s_{\beta-\alpha}$, while those of the  decay rates of $h \to f\bar{f}$ strongly depends on the type of the Yukawa interactions and the type of fermions. 
For instance, in the Type-I 2HDM, the deviations in the decay rate of $h \to f\bar{f}$ are well described by the tree level scaling factor $\kappa_f$, but 
those of $h \to b\bar{b}$ in the Type-X 2HDM can be quite large depending on values of $M^2$ and $c_{\beta-\alpha}$. 
Using our new scheme, we can input the devations in the decay rates of $h \to ZZ^*$ and $h \to \tau\tau$ from the SM value, which will be measured in future Higgs factories, and then 
we can compare the predictions of the other decay modes and the corresponding measured values.

\bibliography{references}



\end{document}